\documentclass[a4paper,11pt]{article}

\usepackage{graphicx,amssymb,amsmath,color}
\usepackage[colorlinks=true, urlcolor=blue, linkcolor=blue, citecolor=blue, bookmarks, pdftitle={article}, pdfauthor={Jens Kunstmann}]{hyperref}
\usepackage[left=2.5cm,right=2.5cm,top=3cm,bottom=3cm]{geometry}

\usepackage{subfigure} 

\newcommand{\bo}{{$\mathrm{B_2O}$}}
\newcommand{\bbo}{{$\mathrm{B_2O_3}$}}

\newcommand{\ea}{\textit{et al.}}

\title{Thermodynamic stability of Borophene, \bbo\ and other $\mathrm{B_{1-x}O_x}$ sheets}

\author{
Florian M.~Arnold$^{1\dag}$, 
Gotthard Seifert$^{1}$,
Jens Kunstmann$^{1\dag}$
\thanks{
$^1$ Theoretical Chemistry, Department of Chemistry and Food Chemistry, TU Dresden, 01062 Dresden, Germany \newline
$\S$ Correspondence should be addressed to J.K.~(e-mail: jens.kunstmann@tu-dresden.de) or F.M.A.~(e-mail: florian.arnold@tu-dresden.de)
}
}

\date{}

\begin{document}

\maketitle

keywords: 2D materials, borophene, B2O3, B2O, boron-oxygen, oxidation, phase diagram, convex hull

\begin{abstract}
The recent discovery of borophene, a two-dimensional allotrope of boron, raises many questions about its structure and its chemical and physical properties. Boron has a high chemical affinity to oxygen  but little is known about the oxidation behaviour of borophene.
Here we use first principles calculations to study the  phase diagram of free-standing, two-dimensional $\mathrm{B_{1-x}O_x}$ for compositions ranging from $x=0$ to $x=0.6$, which correspond to borophene and \bbo\ sheets, respectively.
Our results indicate that no stable compounds except borophene and \bbo\ sheets exist. 
Intermediate compositions are heterogeneous mixtures of borophene and \bbo.
Other hypothetical crystals such as \bo\ are unstable and some of them were found to undergo spontaneous disproportionation into borophene and \bbo. 
It is also shown that oxidizing borophene inside the flakes is thermodynamically unfavorable over forming \bbo\ at the edges.
All findings can be rationalized by oxygen's preference of two-fold coordination which is incompatible with higher in-plane coordination numbers preferred by boron.
These results agree well with recent experiments and pave the way to understand the process of oxidation of borophene and other two-dimensional materials.
\end{abstract}

\section*{Introduction}
\begin{figure}[tb]
\centering
\begin{tabular}{llll}
(a) $(0, 0)$ &
(b) $(0.6, 0)$ &
(c) $(1/3, 0.37)$ &
(d) $(1/3, 0.18)$ \\
\includegraphics[width=0.22\columnwidth]{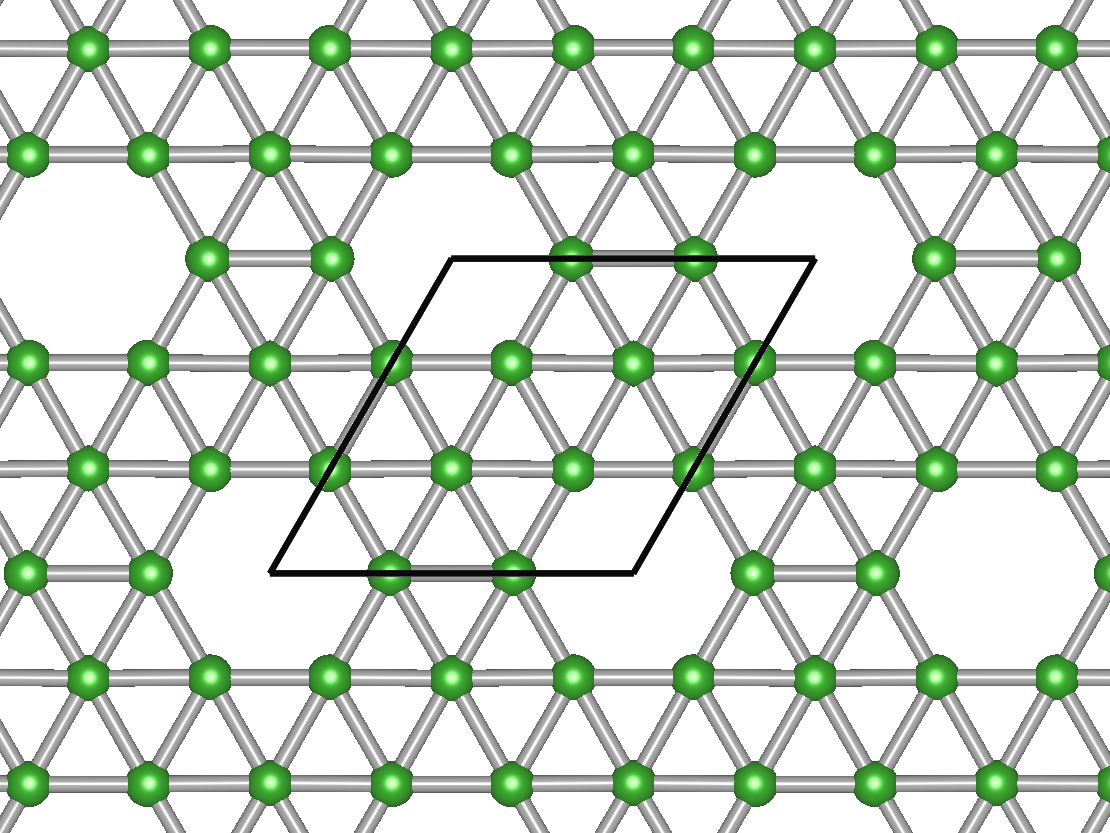}&
\includegraphics[width=0.22\columnwidth]{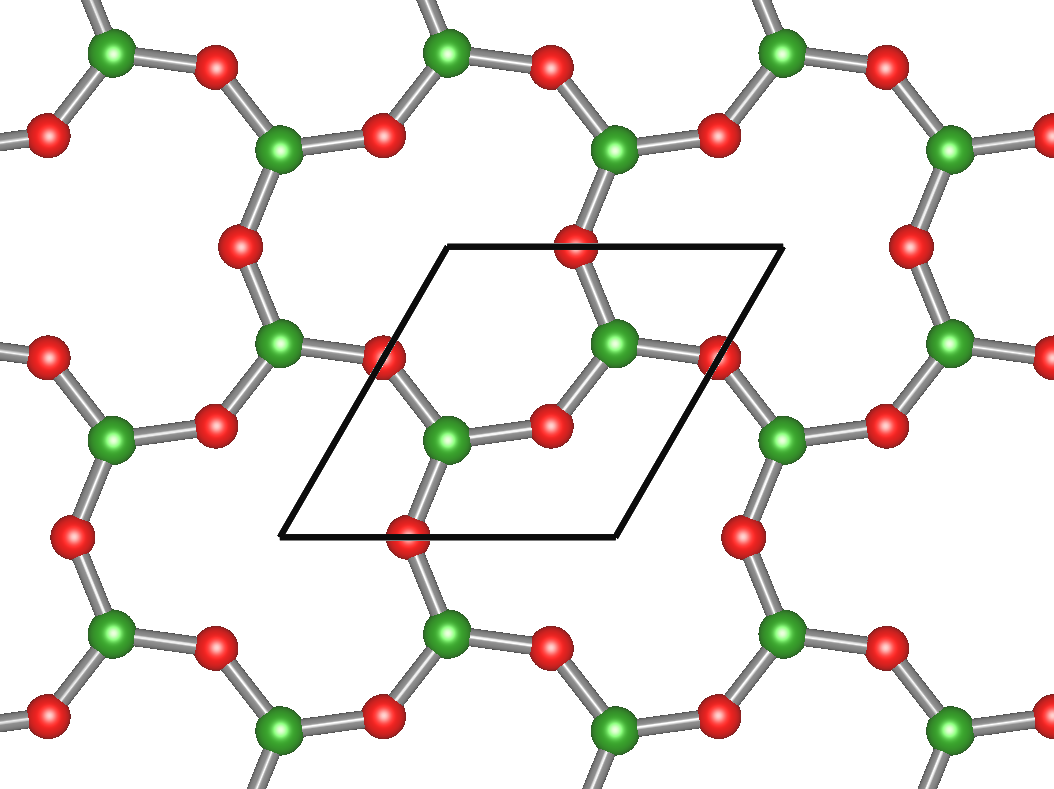}&
\includegraphics[width=0.22\columnwidth]{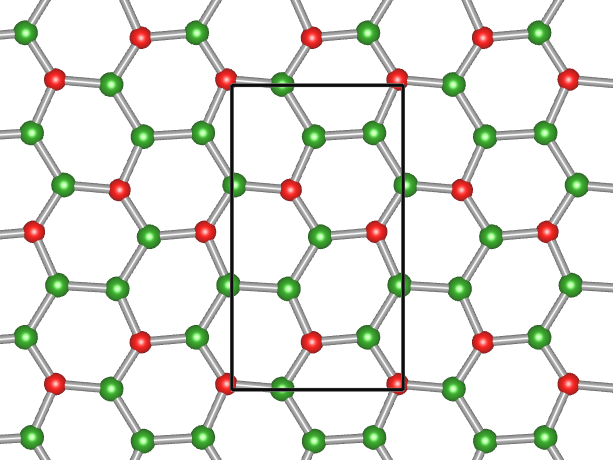}&
\includegraphics[width=0.22\columnwidth]{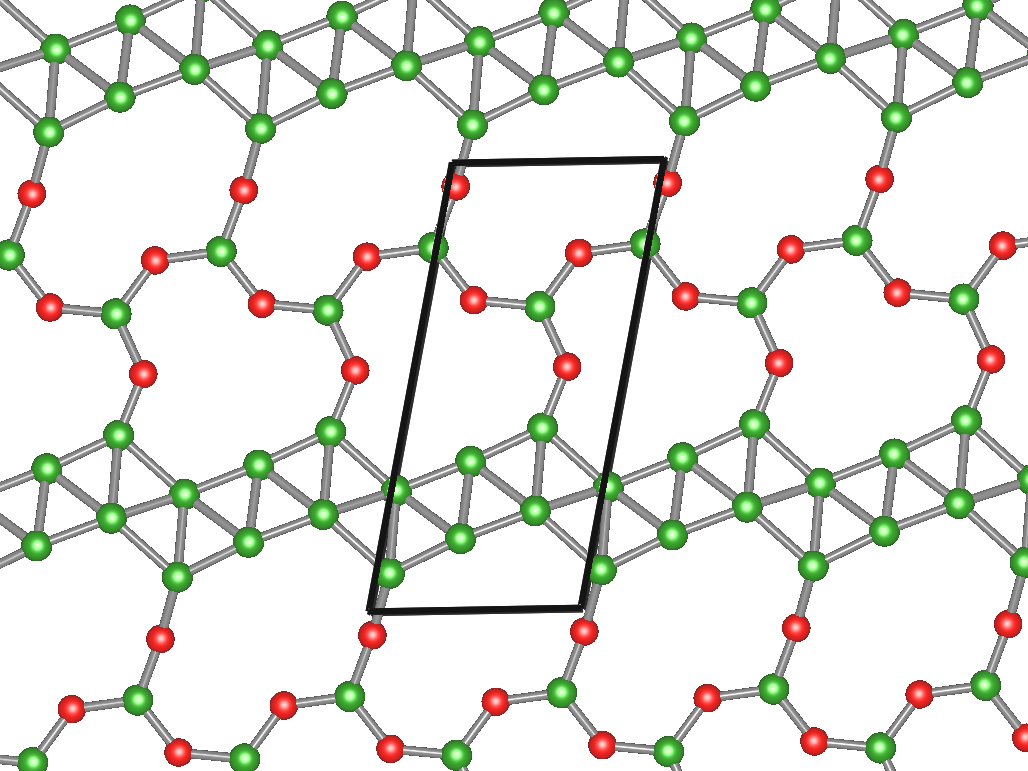}\\
(e) $(0.37, 0.18)$&
(f) $(0.29, 0.18)$ &
(g) $(0.24, 0.20)$ &
(h) $(0.02, 0.04)$\\
\includegraphics[width=0.22\columnwidth]{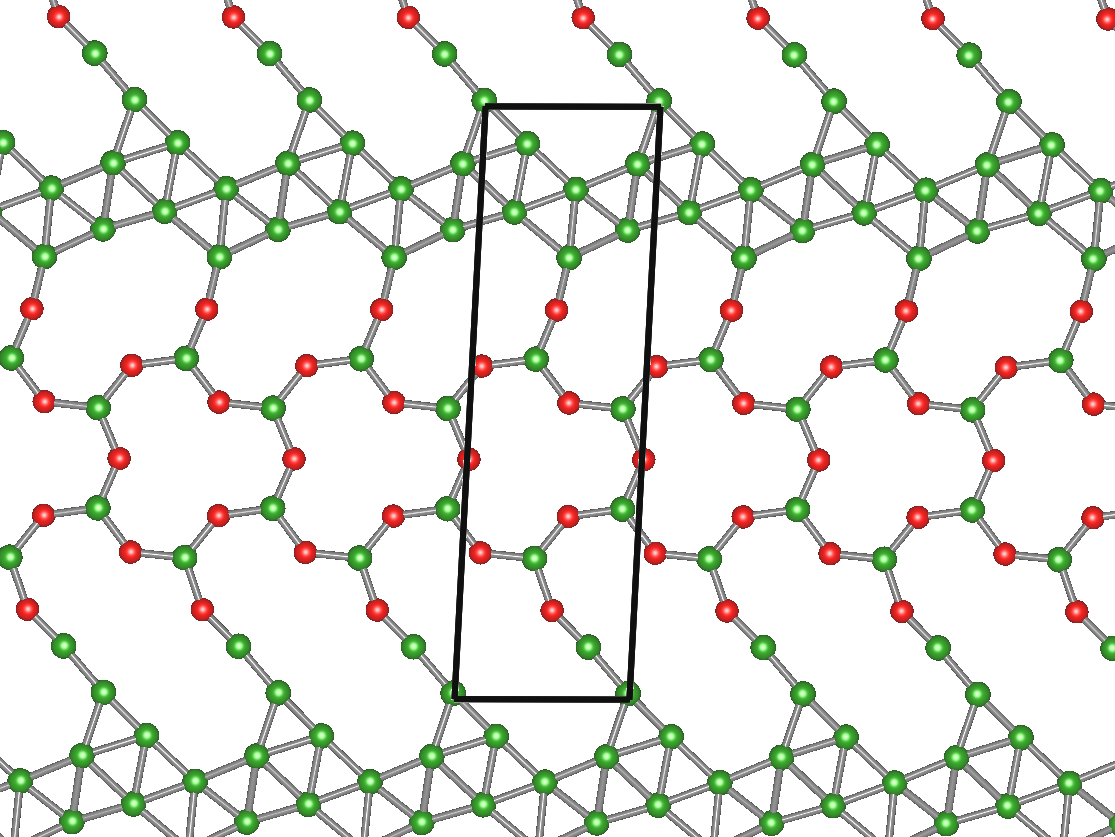}&
\includegraphics[width=0.22\columnwidth]{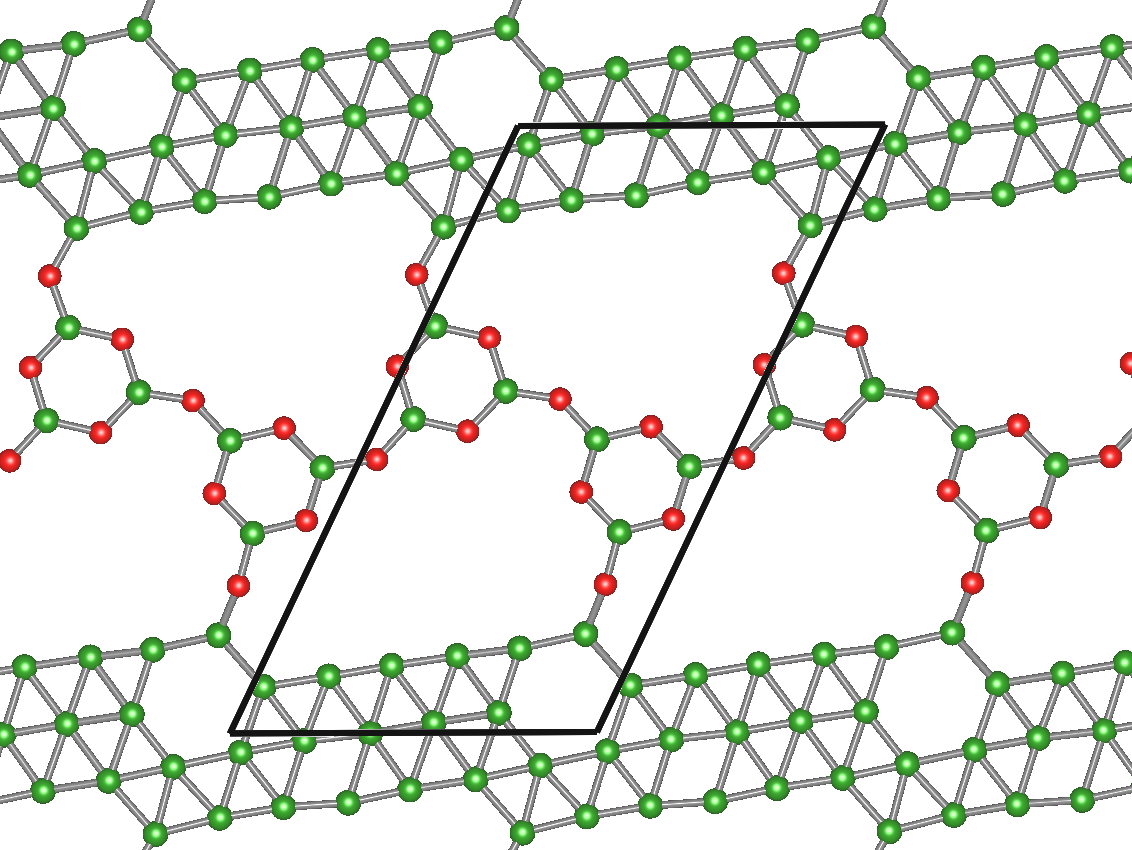}&
\includegraphics[width=0.22\columnwidth]{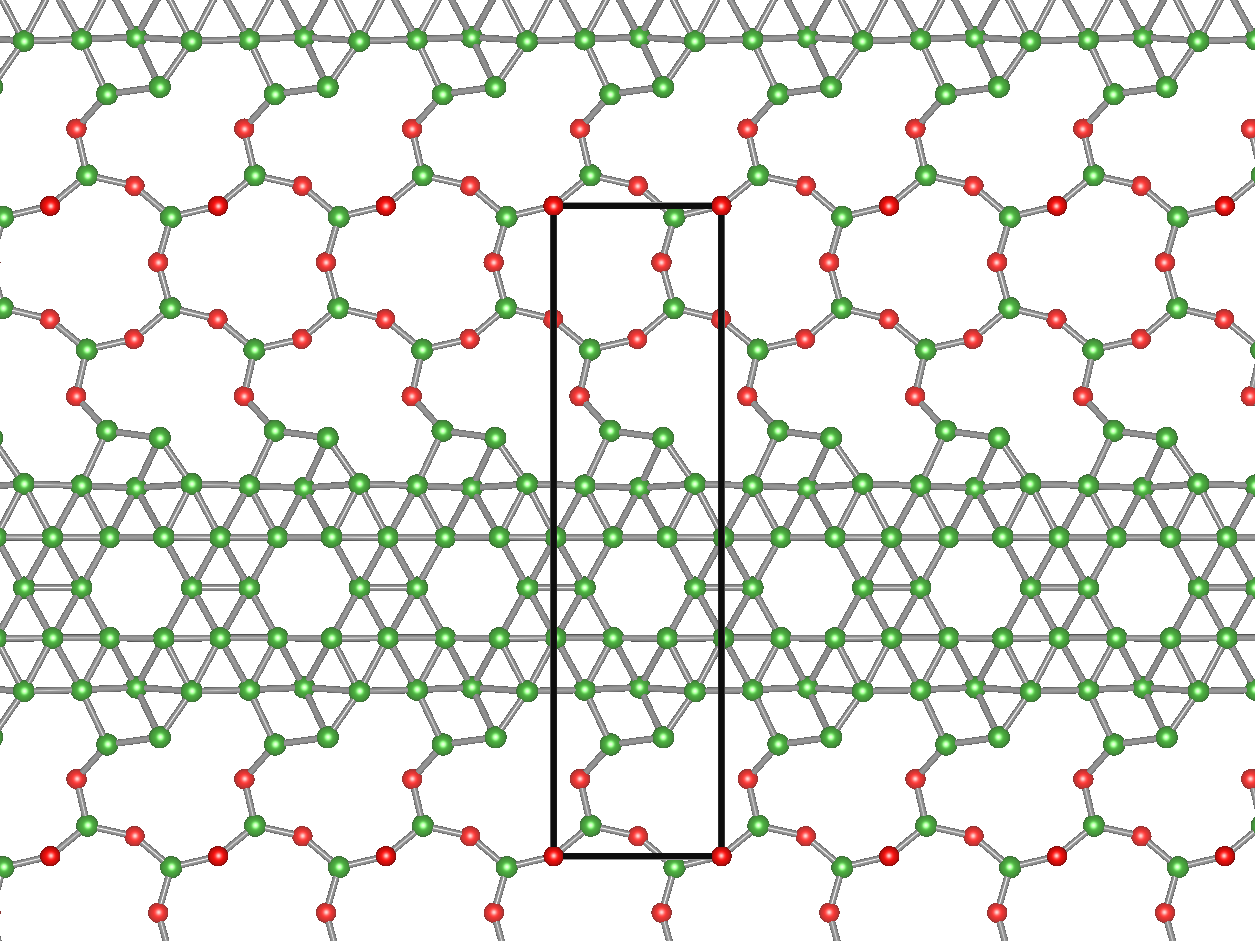}&
\includegraphics[width=0.22\columnwidth]{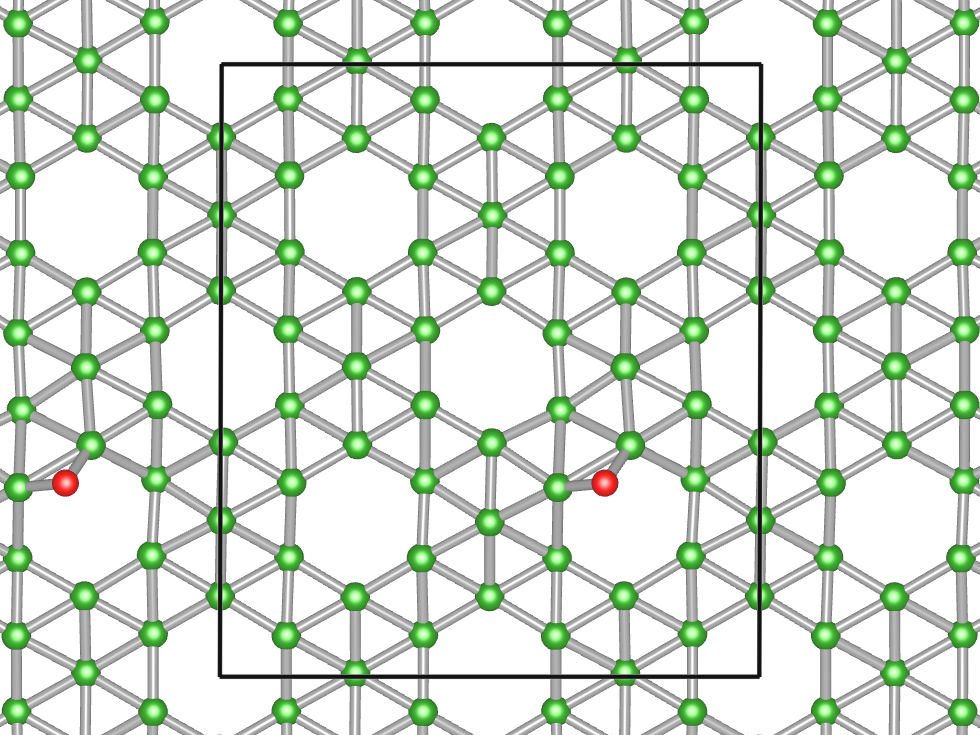}\\
&&&
\includegraphics[width=0.22\columnwidth]{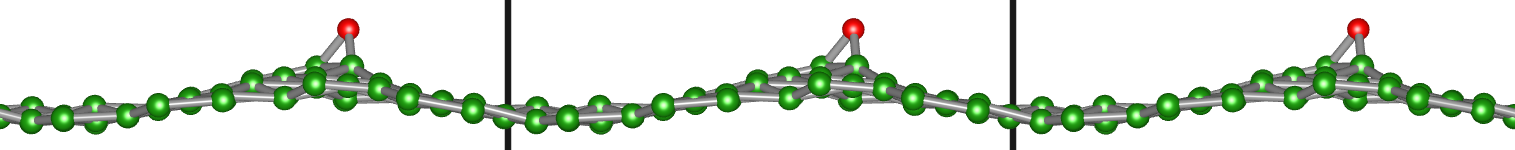}\\
\end{tabular}
\caption{The structure of some two-dimensional $\mathrm{B_{1-x}O_x}$ compounds.
The tuples in the headlines correspond to $(x,E_{\mathrm{mix}})$, where $x$ is the oxygen fraction and  $E_{\mathrm{mix}}$ is the mixing energy in eV/atom (see Eqn.~\ref{Eqn_Emix} and Fig.~\ref{Fig_phasediagram}). Red and green balls represent oxygen and boron atoms, respectively.
(a) Borophene $\alpha'$-sheet. 
(b) \bbo-sheet, introduced by Ferlat \ea\ \cite{Ferlat2012}.
(c) A \bo-sheet with a honeycomb-like structure.
(d)/(e) Systems that spontaneously separated into regions consisting of borophene and \bbo\ during a structural optimization. 
(f)/(g) Larger systems that were constructed to be separated.
(h) Oxygen on borophene, exhibiting divalent out-of-plane bonding, as discernible from the top view (top) and side view (bottom).
}
\label{Fig_structures}
\end{figure}

The investigation and development of two-dimensional (2D) materials is currently a focus of research worldwide
\cite{Novoselov2016,Balleste2011}.
This class of materials was recently extended by another representative when two research teams were able to grow a boron monolayer on a silver surface in ultra-high vacuum \cite{Mannix2015,Feng2016}. The experimental discovery of this "borophene" was anticipated by various theoretical predictions ranging back to the mid-1990s  \cite{Boustani1997c,Kunstmann2006,Tang2007,Wu2012}. Analogous to the existence of many boron bulk phases, free-standing borophene exhibits a pronounced polymorphism \cite{Tang2007,Penev2012}; however it is lifted in the vicinity of a metal surface which enables the growth of specific borophene crystals \cite{Liu2013b,Mannix2018}. A prototypical borophene (the $\alpha'$-sheet) is shown in Fig.~\ref{Fig_structures}(a).

Boron is also known to have a high chemical affinity to oxygen and therefore boron-rich (nano)-structures can only be prepared under inert conditions or vacuum; so is borophene. 
For bulk phases the boron-oxygen system is well studied and several works on the phase diagram exist \cite{Nieto-Sanz2004,Solozhenko2008,Dong2018}. Thermodynamically stable are the pure "icosahredal"  boron phases in their various forms \cite{Shirai2017a}, boron suboxide $\mathrm{B_6O}$ \cite{Hubert1998} and boron trioxide \bbo\ that is a vitreous phase under ambient conditions \cite{Ferlat2012} but can be crystalline when synthesized under pressure \cite{Gurr1970,Prewitt1968,Dong2018}. 
Sometimes boron monoxide \bo\ \cite{Hall1965,Endo1987} is also considered but its stability and existence is strongly debated \cite{Grumbach1995,Hubert1996,Nieto-Sanz2004,Solozhenko2008}. 
However, as the bonding in borophene is different from the bonding in the bulk phases, we can expect it to have different chemical properties, which are largely unexplored so far. 
Feng \ea\ exposed borophene flakes to different oxygen concentrations and found that they tend to oxidize from the edges, while boron atoms inside the flakes are relatively inert to oxidation \cite{Feng2016}. We will come back to this point in the discussion below.
In the literature a variety of 2D boron-oxygen structures were studied.
Two-dimensional variants of \bbo\ were proposed by Ferlat \ea\ \cite{Ferlat2012}, with building blocks formed by planar $\mathrm{BO_3}$-units (see Fig.~\ref{Fig_structures}(b)) or boroxo-rings (see Supplementary Material).
Similar \bbo\ sheets where obtained as part of $\mathrm{A_3HB_4S_2O_{14}}$ (A=Rb,Cs) crystals
by Daub \ea\ \cite{Daub2015}.
Hexagonal \bo\ is a (debated) layered van der Waals crystal
\cite{Hall1965} and  
multiple hypothetical structural models for  monolayers were previously considerd \cite{Zhang2002,Zhong2019}. One possible \bo\ monolayer model is shown in Fig.~\ref{Fig_structures}(c).
Several authors theoretically studied the adsorption of oxygen on the buckled triangular borophene \cite{Lherbier2016,Alvarez-Quiceno2017,Guo2017,He2019} or related nanoribbons \cite{Kistanov2019}.
However, structures related to the buckled triangular structure are not thermodynamically favorable neither as stand-alone system nor when placed on a metal surface \cite{Mannix2018} and angle-resolved photoemission spectroscopy of the experimentally realized borophenes ($\beta_{12}$ and $\chi_3$ sheets) are also highly consistent with non-triangular borophenes  \cite{Feng2017,Feng2018}.
A different theoretical study was performed by Luo \ea\ who considered oxygen adsorption, dissociation and diffusion  on $\chi_3$ borophene on Ag(111) \cite{Luo2018}. They find that oxygen is not incorporated into the borophene layer (which mostly remains structurally intact) but it rather adsorbs on top of it.
Sheets of varying composition, where oxygen is incorporated into the boron plane, were studied by Zhang \ea\ \cite{Zhang2017g} and Lin \ea\  \cite{Lin2018a} considered B-O sheets that are inspired by the structure of planar \bbo.
Kambe \ea\ obtained a layered $\mathrm{B_{4.27}KO_3}$ compound where planar, anionic B-O layers are stabilized by potassium cations between the layers \cite{Kambe2019}.

To systematize and unify the view on these various structures we study 149 boron-oxygen layers by first principles calculations and use them to construct the convex hull of the 2D boron-oxygen system. 
The convex hull is related to the phase diagram and allows to decide which of these many systems are thermodynamically favorable and are likely to be realized experimentally. This also helps to understand the process of oxidation of borophene.
Our findings shed a new light on the boron-oxygen binary system, put the previous literature into context and agree well with experimental reports.

\section*{Computational methods}
The first principles calculations of 2D $\mathrm{B_{1-x}O_x}$ structures were performed with the density functional theory (DFT) code SIESTA, version 3.2 \cite{SIESTA}, using norm-conserving Troullier-Martins pseudopotentials \cite{Troullier1991} as provided on the SIESTA homepage as part of the Abinit's pseudo database.
Electronic correlations were treated using the Perdew-Burke-Ernzerhof (PBE) exchange-correlation functional \cite{PBE} within the generalized gradient approximation. Calculations were carried out using a default DZP basis set and an energy cutoff of  250 Ry for the real-space grid. The unit cells were built with 20 {\AA} separation between replicas in the perpendicular direction to achieve negligible interactions.
The $k$-space integrations were carried out using the Monkhorst-Pack scheme \cite{Monkhorst1976}, employing a Fermi smearing of 300 K.  The $k$-grids were  converged for every considered structure such that energy changes were smaller than 1 meV/atom. 
Atomic coordinates as well as lattice parameters were relaxed using the conjugated gradient method with force tolerance 0.01 eV/{\AA} and stress tolerance 0.1 GPa.

\begin{figure}[tb]
\centering
\includegraphics[width=\columnwidth]{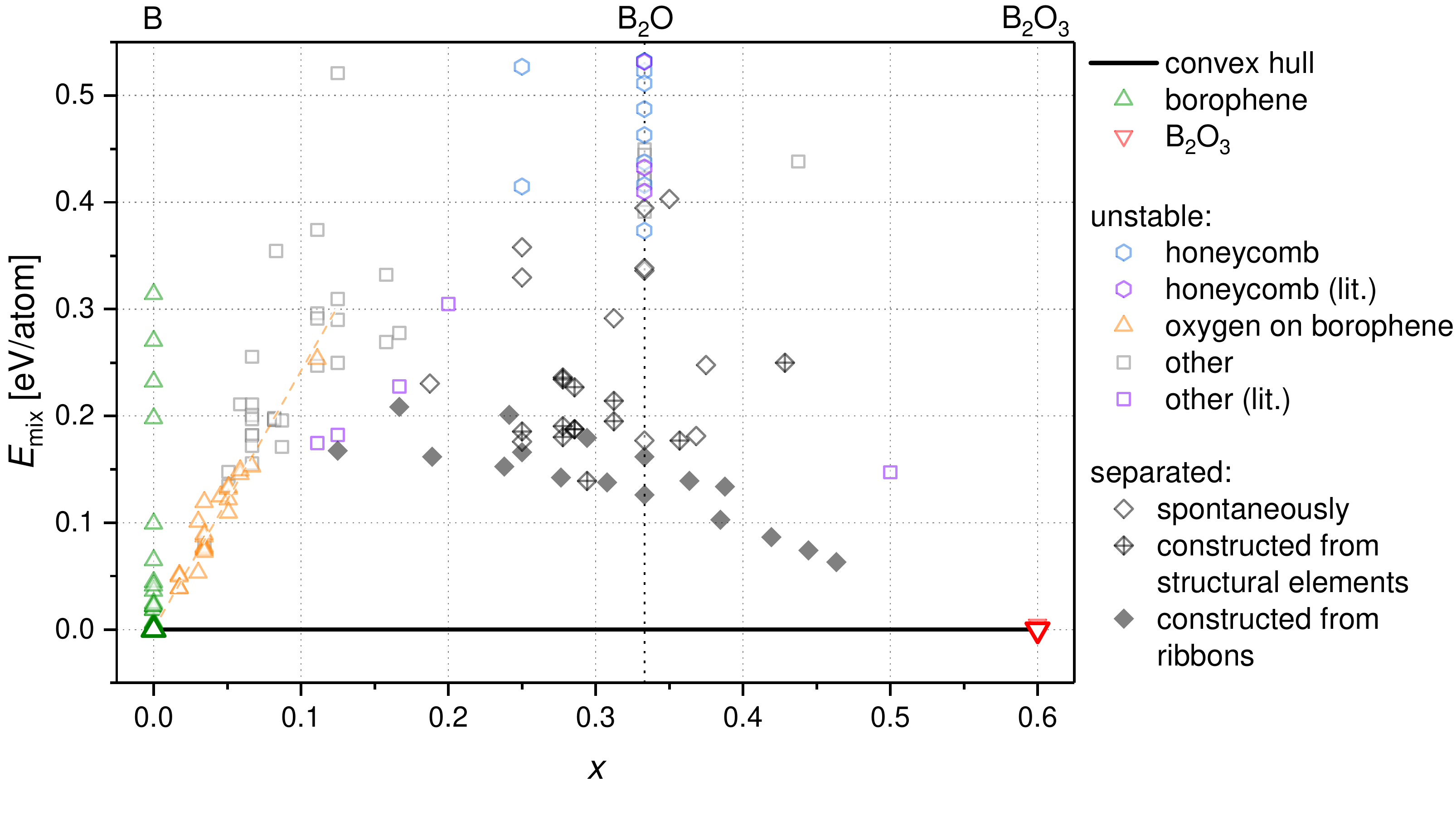}
\caption{Scatter plot of mixing energies $E_{\mathrm{mix}}$ (see Eqn.~\ref{Eqn_Emix}) of free-standing, two-dimensional $\mathrm{B_{1-x}O_x}$ for different oxygen fractions $x$. 
The set of thermodynamically stable compounds in the diagram forms the convex hull. Here it is defined by the tie line connecting the borophene $\alpha'$-sheet for $x=0$ (see Fig.~\ref{Fig_structures}(a)) and a \bbo\ sheet for $x=0.6$ (see Fig.~\ref{Fig_structures}(b)). No other points are on the convex hull; consequently intermediate compositions are heterogeneous mixtures of borophene and \bbo. 
Other data points  with positive mixing energies represent thermodynamically unstable systems. These include  decorations of the honeycomb-lattice (hexagons), oxygen on borophene (orange triangles), irregular structures (squares) and structures previously reported in the literature (purple symbols) \cite{Zhang2002,Zhang2017g,Lin2018a} .
Diamonds indicate heterogeneous mixtures of borophene and \bbo\ in finite-size unit cells.
Empty, gray diamonds indicate systems that spontaneously separated into borophene and \bbo\ during a structural optimization run.
}
\label{Fig_phasediagram}
\end{figure}

\section*{Results and Discussion}
In order to construct the convex hull of free-standing, two-dimensional allotropes of boron and oxygen we considered a total number of 149 $\mathrm{B_{1-x}O_x}$ structures of varying composition $x$ and symmetry. The geometry, energy and compositions of all systems are given in the Supplementary Material. 
To compare the different structures we define the mixing energy 
\begin{equation}
E_{\mathrm{mix}}(y)=E(\mathrm{B_{1-x}O_x})-(1-y)\cdot E(\mathrm{B})-y\cdot E(\mathrm{B_2O_3})
\label{Eqn_Emix}
\end{equation}
where $E(a)$ is the DFT total energy per atom of structure $a$, $E(\mathrm{B})= -79.64$ eV/atom 
is the energy of the borophene $\alpha'$-sheet \cite{Tang2007,Wu2012} and $E(\mathrm{B_2O_3})=-301,33$ eV/atom is the energy of a \bbo\ sheet \cite{Ferlat2012}.
The relative oxygen fraction is given by $y={x}/0.6$; this definition implies that our maximum composition is $x=0.6$ for \bbo.
The mixing energy $E_{\mathrm{mix}}$ can be considered as Gibbs free formation energy $G(x,p,T)$ at $p=0$ Pa and $T=0$ K for the fictitious reaction $(1-y) \ \mathrm{B} + y \ \mathrm{B_2O_3} \rightarrow \mathrm{B_{1-x}O_x}$. 
The results are shown in the scatter plot in Fig.~\ref{Fig_phasediagram}.

The borophene structures on the left-hand side of the plot (green triangles) are taken from various publications \cite{Feng2016,Tang2007,Kunstmann2006,DH,Wu2012}. The rich polymorphism of boron is discernible by the large number of data points. In agreement with the literature we also find that (for the PBE functional) the $\alpha'$-sheet (see Fig.~\ref{Fig_structures}(a)) is the lowest energy borophene for free-standing systems  \cite{Tang2007,Penev2012}. It was therefore chosen as reference structure for calculating $E_{\mathrm{mix}}$.

On the right-hand side of the figure (red triangles) we find another set of very stable structures - the \bbo\ sheets proposed by Ferlat \ea\ \cite{Ferlat2012}. Despite their different structure the two are nearly degenerate in energy. Figure \ref{Fig_structures}(b) shows the \bbo\ sheet with planar $\mathrm{BO_3}$-units that we use as second reference to define $E_{\mathrm{mix}}$. 

The hexagons for a composition of $x=1/3$ in Fig.~\ref{Fig_phasediagram}  correspond to crystalline \bo\ structures. The initial structures were constructed using a honeycomb lattice (for details see Supplementary Material). We think this is a reasonable starting point as formally \bo\ is  isoelectronic to graphene \cite{Hall1965}. We constructed 21 new structures of \bo\ in addition to the three proposed by Zhang \ea\ \cite{Zhang2002} (indicated by purple hexagons). 
The honeycomb-like structure with the lowest mixing energy of 0.37 eV/atom is shown  in Fig.~\ref{Fig_structures}(c). To our knowledge it was not reported before.
However, all these honeycomb-like \bo\ systems still have rather high, positive mixing energies, which indicates that they are thermodynamically unfavorable. 
Besides the many \bo\ realizations on a honeycomb lattice 
we also find systems with rather irregular, non-hexagonal geometry (gray squares) and three structures with a particularly low energy (empty, gray diamonds).
One of them is shown in Fig.~\ref{Fig_structures}(d) and its mixing energy is 0.18 eV/atom;  much lower than the honeycomb-like structures.
These three systems have an inhomogeneous distribution of boron and structurally transformed into regions of pure boron and regions consisting of boron and oxygen  which exhibit $\mathrm{BO_3}$ units, the typical structural unit of \bbo. 
These findings suggest that \bo\ is unstable with respect to spontaneous separation into boron and \bbo\ as described by
\begin{equation*}
\mathrm{3\,B_2O\rightarrow4\,B+B_2O_3}.
\end{equation*}
This is a disproportionation reaction, in which \bo\ (where B is in an intermediate oxidation state of +1) converts to two different compounds, borophen (with a lower oxidation state of 0) and \bbo\ (with a higher  oxidation state of +3). 
A chemically equivalent example is the reaction $\mathrm{3\,Al_2O\rightarrow4\,Al+Al_2O_3}$, which is discussed  in textbooks \cite{Remy70}.
As the disproportionation/separation of \bo\ was found as a result of a simple structural optimization, it means that it occurs without energy barrier. This indicates that \bo\ is not metastable (i.e.~protected by energy barriers) but unstable. It should be mentioned that we only find such a separation if the distribution of oxygen and boron is inhomogeneous in the initial structures. 

To further investigate the apparent tendency of honeycomb-like B-O to disproportionate into heterogeneous mixtures of borophene and \bbo, we extended the study to a broader stoichiometric range.
Structures were again constructed  by placing atoms on the sites of a honeycomb lattice, but now the B:O ratio was varied and boron and oxygen were deliberately distributed inhomogeneously. During the geometry optimization some of the systems remained honeycomb-like (hexagons in Fig.~\ref{Fig_phasediagram}), some transformed into irregular structures (squares), but most of the systems spontaneously separated into borophene and \bbo\ (empty diamonds). An example of the last category is shown in Fig.~\ref{Fig_structures}(e), which corresponds to $x=0.37$ and $E_{\mathrm{mix}} = 0.18$ eV/atom. The strong structural changes that are associated with the disproportionation are illustrated in the Supplementary Material, where all initial and final structures are given.
Thus spontaneous disproportionation/separation occurs over a broad stoichiometric range and separated systems tend to have low mixing energies. 
This is a very strong hint that the corresponding layered \bo\ bulk phase, reported by Hall \ea\ \cite{Hall1965}, is unlikely to exist, as well.
Our findings are also in good agreement with those of Grumbach \ea\ \cite{Grumbach1995}, who found hints for disproportionation in diamond-like \bo\ \cite{Endo1987}, very similar to the one described here. From this point of view, our results can be seen as a possible explanation why there are no conclusive reports about the existence of bulk \bo\ \cite{Hubert1996,Nieto-Sanz2004,Solozhenko2008}.

To lower $E_{\mathrm{mix}}$ further, structures that are already separated in their initial geometry were studied. The typical structural elements used for this approach were triangular units for boron and $\mathrm{BO_3}$-units and boroxo-rings for \bbo. These were connected to form areas of pure boron linked to areas of \bbo.
An example from this set of structures is shown in Fig.~\ref{Fig_structures}(f).
The energies are shown as crossed, gray diamonds in Fig.~\ref{Fig_phasediagram}.
It is obvious that $E_{\mathrm{mix}}$ is significantly lower for such systems. 
The last approach to construct separated systems was by using boron and \bbo\ ribbons (for details see the Supplementary Material).  An example for these structures is shown in Fig.~\ref{Fig_structures}(g).
These structures show the overall lowest mixing energies of all studied sheets (filled, gray diamonds in Fig.~\ref{Fig_phasediagram}).
We also added the energy of the structures studied by Zhang \ea\ \cite{Zhang2017g} and Lin \ea\  \cite{Lin2018a} to Fig.~\ref{Fig_phasediagram} (purple symbols). 
Their mixing energy is significantly higher than the ones of the heterogeneous mixtures, indicating that they are thermodynamically unfavorable. 
All the described approaches yield the same result for the considered range of compositions - the lowest energy structures are heterogeneous mixtures of borophene and \bbo.

To generate systems with with small oxygen fractions 50 structures were constructed by adding oxygen atoms to different borophene sheets ($\alpha$, $\alpha_1$, $\beta$, $\beta_1$, $\chi_3$ following the notation by Wu \ea\ \cite{Wu2012}). 
We found that for $x<0.05$ (orange triangles in Fig.~\ref{Fig_phasediagram}) the systems are mostly characterized by divalent oxygen atoms binding out-of-plane to sites near the hexagonal holes of the borophene sheets. The sheets themselves remain intact and are structurally modified only in the vicinity of the O atom. Figure \ref{Fig_structures}(h) shows one example of such a system. These findings agree well with similar results by Luo \ea\ \cite{Luo2018}. 
For $x \ge 0.05$ oxygen tends to be incorporated in-plane and the structure is distorted (indicated by squares in Fig.~\ref{Fig_phasediagram}). 
The data points in Fig.~\ref{Fig_phasediagram} for small $x$  rise linearly and the slope ($\mu_o = 2.53$ eV/atom) is the chemical potential. Mind that $\mu_o$ is defined relative to borophene and \bbo. So the positive value means that oxidizing borophene inside the flakes is thermodynamically unfavorable over forming heterogeneous mixtures of borophene and \bbo. 
This is in full agreement with the experimental observation by Feng \ea, who found that borophene flakes tend to oxidize from the edges (forming \bbo), while boron atoms inside the flakes are relatively inert to oxidation \cite{Feng2016}. 
Further support for these results comes from our finding that the edge energy of a \bbo\ ribbon is only 0.25 eV per edge atom, while the edge energy of a borophene ribbon is 1.62 eV/atom, i.e., more than 6 times bigger (see Supplementary Material). We can assume that the edge energy of a flake is similar to the one of a ribbon. 
Then the oxidation of a borophene flake from the edges (by forming \bbo) not only allows the system to minimize the mixing energy, that represents the bulk of the flake, but also to reduce the edge energy.

From a chemical point of view all our results are easy to explain. In honeycomb-like \bo-structures oxygen atoms are 3-fold coordinated, incorporated into borophene layers, oxygen would be 6-fold coordinated. However it is a well-known fact that oxygen prefers coordination numbers lower than 3 and this is only possible in \bbo, where its coordination number is two.
And indeed strong structural reorganisations, leading to the formation of \bbo\ areas, are mostly observed for systems where oxygen is over-coordinated in the initial structure. 

Considering all data points in Fig.~\ref{Fig_phasediagram}, we are now in the position to determine the convex hull. It is defined by the set of thermodynamically stable compounds.
For the 2D $\mathrm{B_{1-x}O_x}$ binary system and compositions $0 \le x \le 0.6$ the convex hull is equivalent to the tie line connecting the borophene $\alpha'$-sheet for $x=0$ and the \bbo\ sheet for $x=0.6$. No other points are on the convex hull and 
the structures with positive mixing energies represent thermodynamically unstable systems, which include honeycomb-like \bo\ \cite{Hall1965,Zhang2002} or other previously considered structures \cite{Zhong2019,Zhang2017g,Lin2018a}.
The gray diamonds in Fig.~\ref{Fig_phasediagram} lie above the convex hull. They represent heterogeneous mixtures of borophene and \bbo\ in finite size unit cells. In these cells relatively small areas of borophene and \bbo\ are connected by interfaces, that occupy a considerable part of the total cell area. Thus the interface energy contributes strongly to the mixing energy. 
In the discussion above we showed that the mixing energy can easily be reduced by choosing larger unit cells with smaller interface-to-area ratios. In the thermodynamic limit the areas of the individual phases are large and the interface-to-area ratio tends to zero. In this case all data points with gray diamonds would be exactly on the convex hull, that is the Gibbs free energy of the compounds at zero temperature and zero pressure. This demonstrates that intermediate compositions of the 2D $\mathrm{B_{1-x}O_x}$ binary system are heterogeneous mixtures of borophene and \bbo.

\section*{Conclusion}
In this work we used first principles calculations to study the compositional phase diagram of free-standing, two-dimensional $\mathrm{B_{1-x}O_x}$ crystals for compositions ranging from $x=0$ to $x=0.6$, which correspond to borophene and \bbo\ sheets, respectively.
Our results show that the convex hull of the phase diagram is defined by borophene and \bbo\ and no other  phases. 
Intermediate compositions are thus heterogeneous mixtures of these two compounds.
Hypothetical crystals with intermediate compositions such as \bo\ ($x=1/3$) are unstable, because their energies are significantly above the convex hull and some of them were found to undergo spontaneous disproportionation into borophene and \bbo. This could explain why in the literature there are no conclusive reports  about the existence of bulk \bo. 
It is also found that oxidizing borophene inside the flakes is thermodynamically unfavorable over forming heterogeneous mixtures of borophene and \bbo, fully consistent with previous reports on the tendency of borophene to oxidize from its edges (forming \bbo) rather than from the inside \cite{Feng2016}. 
All findings can be rationalized by oxygen's preference of two-fold coordination which is incompatible with higher in-plane coordination numbers favored by boron.

These results are an important step forward to understand the oxidation behavior of 2D boron, which is crucial for its practical use in potential future electronic, optical or chemical applications. 
For future investigations it would be interesting to reconsider the compositional phase diagram on metal surfaces as in the experiment, which might be simplified by studying systems as a function of global charge transfer.

\subsection*{Acknowledgements}
The work is financially supported by the German Research Foundation (DFG) under grant numbers SE 651/45-1. 
F.A. is financially supported by "Deutschlandstipendium".
Computational resources for this project were provided by ZIH Dresden under project "nano-10". We thank Igor Baburin (TU Dresden) for helpful discussions.

\subsection*{Additional information}
Supplementary material is available for this paper upon request by the authors. It contains the energies, compositions and structural illustration for all considered systems.

\bibliographystyle{prsty-con-titoli}
\bibliography{../Literatur,../jens}

\begin{thebibliography}{10}

\bibitem{Novoselov2016}
K.~S. Novoselov, A. Mishchenko, A. Carvalho, and A.~H. {Castro Neto}, {\em {2D
  materials and van der Waals heterostructures}}, Science.
  \textbf{353},  aac9439  (2016).

\bibitem{Balleste2011}
R. Mas-Balleste, C. Gomez-Navarro, J. Gomez-Herrero, and F. Zamora, {\em 2D
  materials: to graphene and beyond}, Nanoscale \textbf{3},  20  (2011).

\bibitem{Mannix2015}
A.~J. Mannix \textit{et~al.}, {\em Synthesis of borophenes: Anisotropic,
  two-dimensional boron polymorphs}, Science \textbf{350},  1513  (2015).

\bibitem{Feng2016}
B. Feng \textit{et~al.}, {\em Experimental realization of two-dimensional boron
  sheets}, Nature Chem. \textbf{8},  563  (2016).

\bibitem{Boustani1997c}
I. Boustani, {\em {New quasi-planar surfaces of bare boron}}, Surf. Sci.
  \textbf{370},  355  (1997).

\bibitem{Kunstmann2006}
J. Kunstmann and A. Quandt, {\em Broad boron sheets and boron nanotubes: an ab
  initio study of structural, electronic, and mechanical properties}, Phys.
  Rev. B \textbf{74},  035413  (2006).

\bibitem{Tang2007}
H. Tang and S. Ismail-Beigi, {\em Novel precursors for boron nanotubes: the
  competition of two-center and three-center bonding in boron sheets}, Phys.
  Rev. Lett. \textbf{99},  115501  (2007).

\bibitem{Wu2012}
X. Wu \textit{et~al.}, {\em Two-dimensional boron monolayer sheets}, ACS Nano
  \textbf{6},  7443  (2012).

\bibitem{Penev2012}
E.~S. Penev, S. Bhowmick, A. Sadrzadeh, and B.~I. Yakobson, {\em Polymorphism
  of two-dimensional boron}, Nano Lett. \textbf{12},  2441  (2012).

\bibitem{Liu2013b}
Y. Liu, E.~S. Penev, and B.~I. Yakobson, {\em {Probing the Synthesis of
  Two-Dimensional Boron by First-Principles Computations}}, Angew. Chemie Int.
  Ed. \textbf{52},  3156  (2013).

\bibitem{Mannix2018}
A.~J. Mannix \textit{et~al.}, {\em {Borophene as a prototype for synthetic 2D
  materials development}}, Nat. Nanotechnol. \textbf{13},  444  (2018).

\bibitem{Nieto-Sanz2004}
D. Nieto-Sanz, P. Loubeyre, W. Crichton, and M. Mezouar, {\em X-ray study of
  the synthesis of boron oxides at high pressure: phase diagram and equation of
  state}, Phys. Rev. B \textbf{70},  214108  (2004).

\bibitem{Solozhenko2008}
V.~L. Solozhenko, O.~O. Kurakevych, V.~Z. Turkevich, and D.~V. Turkevich, {\em
  Phase diagram of the B- B2O3 system at 5 GPa: Experimental and theoretical
  studies}, J. Phys. Chem. B \textbf{112},  6683  (2008).

\bibitem{Dong2018}
H. Dong \textit{et~al.}, {\em {Boron oxides under pressure: Prediction of the
  hardest oxides}}, Phys. Rev. B \textbf{98},  174109  (2018).

\bibitem{Shirai2017a}
K. Shirai, {\em {Phase diagram of boron crystals}}, Jpn. J. Appl. Phys.
  \textbf{56},  05FA06  (2017).

\bibitem{Hubert1998}
H. Hubert \textit{et~al.}, {\em {Icosahedral packing of B12 icosahedra in boron
  suboxide (B6O)}}, Nature \textbf{391},  376  (1998).

\bibitem{Ferlat2012}
G. Ferlat, A.~P. Seitsonen, M. Lazzeri, and F. Mauri, {\em Hidden polymorphs
  drive the vitrification in B2O3}, Nature Mater. \textbf{11},  925  (2012).

\bibitem{Gurr1970}
G. Gurr, P. Montgomery, C. Knutson, and B. Gorres, {\em The crystal structure
  of trigonal diboron trioxide}, Acta Cryst. B \textbf{26},  906  (1970).

\bibitem{Prewitt1968}
C. Prewitt and R. Shannon, {\em Crystal structure of a high-pressure form of
  B2O3}, Acta Cryst. B \textbf{24},  869  (1968).

\bibitem{Hall1965}
H.~T. Hall and L.~A. Compton, {\em Group IV analogs and high pressure, high
  temperature synthesis of B2O}, Inorg. Chem. \textbf{4},  1213  (1965).

\bibitem{Endo1987}
T. Endo, T. Sato, and M. Shimada, {\em High-pressure synthesis of B 2 O with
  diamond-like structure}, J. Mater. Sci. Lett. \textbf{6},
  683  (1987).

\bibitem{Grumbach1995}
M.~P. Grumbach, O.~F. Sankey, and P.~F. McMillan, {\em Properties of B 2 O: An
  unsymmetrical analog of carbon}, Phys. Rev. B \textbf{52},  15807
  (1995).

\bibitem{Hubert1996}
H. Hubert \textit{et~al.},  in {\em Mater. Res. Soc. Symp. - Proc.}, edited by
  R. Bormann \textit{et~al.} (Materials Research Society, , 1996), Vol.~410,
  pp.\ 191--196.

\bibitem{Daub2015}
M. Daub and H. Hillebrecht, {\em Borosulfates Cs2B2S3O13, Rb4B2S4O17, and
  A3HB4S2O14 (A= Rb, Cs)--Crystalline Approximants for Vitreous B2O3?},
  Eur. J. Inorg. Chem. \textbf{2015},  4176  (2015).

\bibitem{Zhang2002}
P. Zhang and V.~H. Crespi, {\em Theory of B 2 O and B e B 2 Nanotubes: New
  Semiconductors and Metals in One Dimension}, Phys. Rev. Lett.
  \textbf{89},  056403  (2002).

\bibitem{Zhong2019}
C. Zhong \textit{et~al.}, {\em {Two-dimensional honeycomb borophene oxide:
  strong anisotropy and nodal loop transformation}}, Nanoscale \textbf{11},
  2468  (2019).

\bibitem{Lherbier2016}
A. Lherbier, A.~R. Botello-M{\'{e}}ndez, and J.-C. Charlier, {\em {Electronic
  and optical properties of pristine and oxidized borophene}}, 2D Mater.
  \textbf{3},  045006  (2016).

\bibitem{Alvarez-Quiceno2017}
J.~C. Alvarez-Quiceno, R.~H. Miwa, G.~M. Dalpian, and A. Fazzio, {\em
  {Oxidation of free-standing and supported borophene}}, 2D Mater. \textbf{4},
  025025  (2017).

\bibitem{Guo2017}
R.-Y. Guo, T. Li, S.-E. Shi, and T.-H. Li, {\em {Oxygen defects formation and
  optical identification in monolayer borophene}}, Mater. Chem. Phys.
  \textbf{198},  346  (2017).

\bibitem{He2019}
Y. He \textit{et~al.}, {\em {Tuning the electronic transport anisotropy in
  borophene via oxidation strategy}}, Sci. China Technol. Sci. \textbf{62},
  799  (2019).

\bibitem{Kistanov2019}
A.~A. Kistanov, S.~K. Khadiullin, S.~V. Dmitriev, and E.~A. Korznikova, {\em
  {Effect of oxygen doping on the stability and band structure of borophene
  nanoribbons}}, Chem. Phys. Lett. \textbf{728},  53  (2019).

\bibitem{Feng2017}
B. Feng \textit{et~al.}, {\em {Dirac Fermions in Borophene}}, Phys. Rev. Lett.
  \textbf{118},  096401  (2017).

\bibitem{Feng2018}
B. Feng \textit{et~al.}, {\em {Discovery of 2D Anisotropic Dirac Cones}}, Adv.
  Mater. \textbf{30},  1704025  (2018).

\bibitem{Luo2018}
W. Luo \textit{et~al.}, {\em {The adsorption and dissociation of oxygen on Ag
  (111) supported chi 3 borophene}}, Phys. B Condens. Matter \textbf{537},  1
  (2018).

\bibitem{Zhang2017g}
R. Zhang, Z. Li, and J. Yang, {\em {Two-Dimensional Stoichiometric Boron Oxides
  as a Versatile Platform for Electronic Structure Engineering}}, J. Phys.
  Chem. Lett. \textbf{8},  4347  (2017).

\bibitem{Lin2018a}
S. Lin \textit{et~al.}, {\em {Porous hexagonal boron oxide monolayer with
  robust wide band gap: A computational study}}, FlatChem \textbf{9},  27
  (2018).

\bibitem{Kambe2019}
T. Kambe \textit{et~al.}, {\em {Solution Phase Mass Synthesis of 2D Atomic
  Layer with Hexagonal Boron Network}}, J. Am. Chem. Soc. \textbf{141},  12984
  (2019).

\bibitem{SIESTA}
J.~M. Soler \textit{et~al.}, {\em The SIESTA method for ab initio order-N
  materials simulation}, J. Phys. Condens. Matter \textbf{14},
  2745  (2002).

\bibitem{Troullier1991}
N. Troullier and J.~L. Martins, {\em Efficient pseudopotentials for plane-wave
  calculations}, Phys. Rev. B \textbf{43},  1993  (1991).

\bibitem{PBE}
J.~P. Perdew, K. Burke, and M. Ernzerhof, {\em Generalized gradient
  approximation made simple}, Phys. Rev. Lett. \textbf{77},  3865
  (1996).

\bibitem{Monkhorst1976}
H.~J. Monkhorst and J.~D. Pack, {\em Special points for Brillouin-zone
  integrations}, Phys. Rev. B \textbf{13},  5188  (1976).

\bibitem{DH}
K.~C. Lau, R. Pati, R. Pandey, and A.~C. Pineda, {\em First-principles study of
  the stability and electronic properties of sheets and nanotubes of elemental
  boron}, Chem. Phys. Lett. \textbf{418},  549  (2006).

\bibitem{Remy70}
H. Remy, {\em Lehrb. der Anorg. Chemie Band 1} (Akademische Verlagsgesellschaft
  Geest {\&} Portig, Leipzig, 1970), Chap.~10.

\end{thebibliography}

\end{document}